\documentclass{PoS}

\usepackage{bm}
\usepackage{amsmath}
\usepackage[square,comma,numbers,sort&compress]{natbib}
\usepackage{color}
\usepackage{graphicx}
\usepackage{bbm}

\newcommand*{\La}{\cal{L}}
\newcommand*{\no}{\noindent}
\newcommand*{\bea}{\begin{eqnarray}}
\newcommand*{\eea}{\end{eqnarray}}
\newcommand*{\be}{\begin{equation}}
\newcommand*{\ee}{\end{equation}}
\newcommand*{\pd}{\partial}
\newcommand*{\pdm}{\pd_{\mu}}
\newcommand*{\pdn}{\pd_{\nu}}

\newcommand*{\pref}[1]{(\ref{#1})}

\newcommand*{\mn}{{\mu\nu}}

\newcommand*{\nn}{\nonumber}

\title{Non-perturbative aspects in a weakly interacting Higgs sector}

\ShortTitle{Non-perturbative aspects in the Higgs sector}

\author{\speaker{Axel Maas}\thanks{Supported by the DFG under grant number MA 3935/5-1.}\\
        E-mail: \email{axelmaas@web.de}}

\author{Tajdar Mufti\thanks{Supported by the DFG graduate school 1523-1 and under grant number MA 3935/5-1.}\\
        E-mail: \email{tajdar.mufti@uni-jena.de}}

\abstract{Just like the weakly interacting QED can support non-perturbative phenomena, like atoms, so can the weak and Higgs interactions. Especially, there are strong field-theoretical arguments that only bound states can be the (quasi-)asymptotic physical degrees of freedom of this sector. After a brief review of these arguments, the 2-point, 3-point and 4-point correlation functions of the Higgs-$W$ system are determined using lattice gauge theory. The results support a conjectured duality between elementary states and bound states for weak Higgs self-interactions. This leads to relations between the bound states and the experimentally observed particles. Interestingly, these may yield pseudo-scalar admixtures at the Higgs energy, and possibly a faint standard-model signal in the channel where a Kaluza-Klein graviton would be expected.}

\FullConference{36th International Conference on High Energy Physics,\\
		July 4-11, 2012\\
		Melbourne, Australia}

\begin{document}

\section{Introduction: The need for non-perturbative phenomena in the Higgs sector}

Weakly interacting gauge theories have a surprisingly rich structure. QED is the paradigmatic example, being the most weakly interacting part of the standard model, aside from some of the Yukawa interactions of the Higgs. Nonetheless, atoms are non-perturbative stable bound states.

Similarly, the weak interactions and the Higgs sector can be expected to provide bound states. The most profound reason is possibly that neither the Higgs nor the $W$ and $Z$ are gauge-invariant states, and thus cannot be physical asymptotic states \cite{Frohlich:1981yi}. Less fundamental, but likely as valid aside from triviality questions, is the phase structure of the pure $W$-Higgs sector considered here \cite{Fradkin:1978dv}: Since the would-be Higgs regime and the would-be confinement regime are continuously connected, the asymptotic state space must be the same, irrespective of the strength of the Higgs self-interaction.

Both arguments imply that bound states, described by gauge-invariant composite operators, are the actual physical degrees of freedom. This raises immediately the question of why then a perturbative description, identifying the Higgs and the $W$ as physical particles, describes the experimental results so well. The reason is a duality \cite{Frohlich:1981yi} between these elementary particles, described in more detail in section \ref{sgd2} and \ref{sgd3}, and the bound states, described in section \ref{sgi}. This has quite profound consequences for experimental signatures, see section \ref{sgi} and \cite{Maas:2012tj}.

\section{Gauge-dependent correlators: Propagators}\label{sgd2}

Here, only the pure $W$-Higgs sector is considered, i.\ e.\ a theory described by the Lagrangian
\bea
{\La}&=&-\frac{1}{4}W_\mn^aW^\mn_a+(D_\mu\phi)^+D^\mu\phi-\gamma(\phi\phi^+)^2-\frac{m^2}{2}\phi\phi^+\label{action}\\
W_\mn^a&=&\pdm W_\nu^a-\pdn W_\mu^a-g f^{abc} W_\mu^b W_\nu^c\nn\\
D_\mu^{ij}&=&\pd_\mu\delta^{ij}-igW_\mu^a\tau^{ij}_a\nn,
\eea
\no with the fundamental complex scalar $\phi$ coupled to su(2) valued gauge fields $W_\mu$ (containing the, without QED, degenerate $W$ and $Z$ gauge bosons) with the field strength tensor $W^a_\mn$, the covariant derivative $D_\mu$, the coupling constants $g$, $\gamma$, and $m$, the Pauli matrices $\tau$, and structure constants $f^{abc}$. To obtain non-perturbative results, lattice simulations can be used \cite{Maas:2012tj,Langguth:1985dr,Evertz:1985fc,Philipsen:1996af}. See \cite{Maas:2012tj} for the technical details of the simulations done here, which were performed at a Higgs mass of 153 GeV.

The propagation of the elementary $W$ and Higgs particles are described by the gauge-dependent and renormalization-scheme-dependent 2-point functions $\langle W_\mu^a W_\nu^b\rangle(x-y)$ and $\langle\phi^{+a}\phi^b\rangle(x-y)$, respectively \cite{Bohm:2001yx}. Thus, it is necessary to fix a gauge to determine them, which will be here the non-aligned \cite{Maas:2012ct} minimal Landau gauge \cite{Maas:2011se}, see again \cite{Maas:2012tj} for implementation and renormalization details.

\begin{figure}
\includegraphics[width=\linewidth]{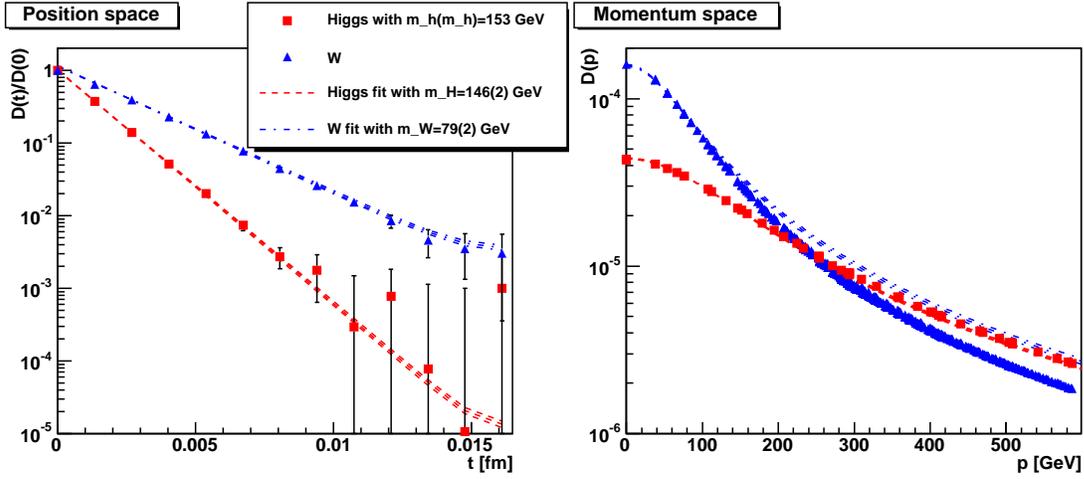}
\caption{\label{fig:2}The two-point functions in position space for zero three-momentum (left panel) and momentum space for various momentum configurations (right panel), compared to tree-level massive particle fits. See \cite{Maas:2012tj} for details of the lattice simulations and parameters. All results from a 24$^4$ lattice at $(147$ GeV)$^{-1}$ lattice spacing (corresponding to an imposed cutoff of roughly 650 GeV), where the scale was set by the $W$ mass.}
\end{figure}

The resulting propagators are shown in figure \ref{fig:2} in both momentum and position space. Note that the mass of the Higgs is scheme-dependent \cite{Bohm:2001yx,Maas:2012tj}. It has been set to a value of 153 GeV for reasons to become apparent in section \ref{sgi}. In position space, both propagators are essentially behaving, within large statistical errors, as simple massive particles. In momentum space, this is also the case for the Higgs particle. However, the $W$ shows a decay at large momentum different from the one of a massive particle. This is expected, since the $W$ is not truly massive, and a faster decay is necessary for the unitarization of cross sections \cite{Bohm:2001yx}. Thus, both the Higgs and the $W$ behave essentially as expected.

\section{Gauge-dependent correlators: Vertices}\label{sgd3}

Besides the particles themselves their interactions are of course highly interesting, as many sensitive tests of the standard model are on electroweak radiative corrections. Though the 4-Higgs coupling, especially its high momentum behavior, would be a decisive information for the question of triviality, it is a challenging problem, due to disconnected contributions, in numerical simulations. Though non-aligned gauges reduce this challenge significantly \cite{Maas:2012ct}, this will require much further development.

\begin{figure}
\includegraphics[width=0.5\linewidth]{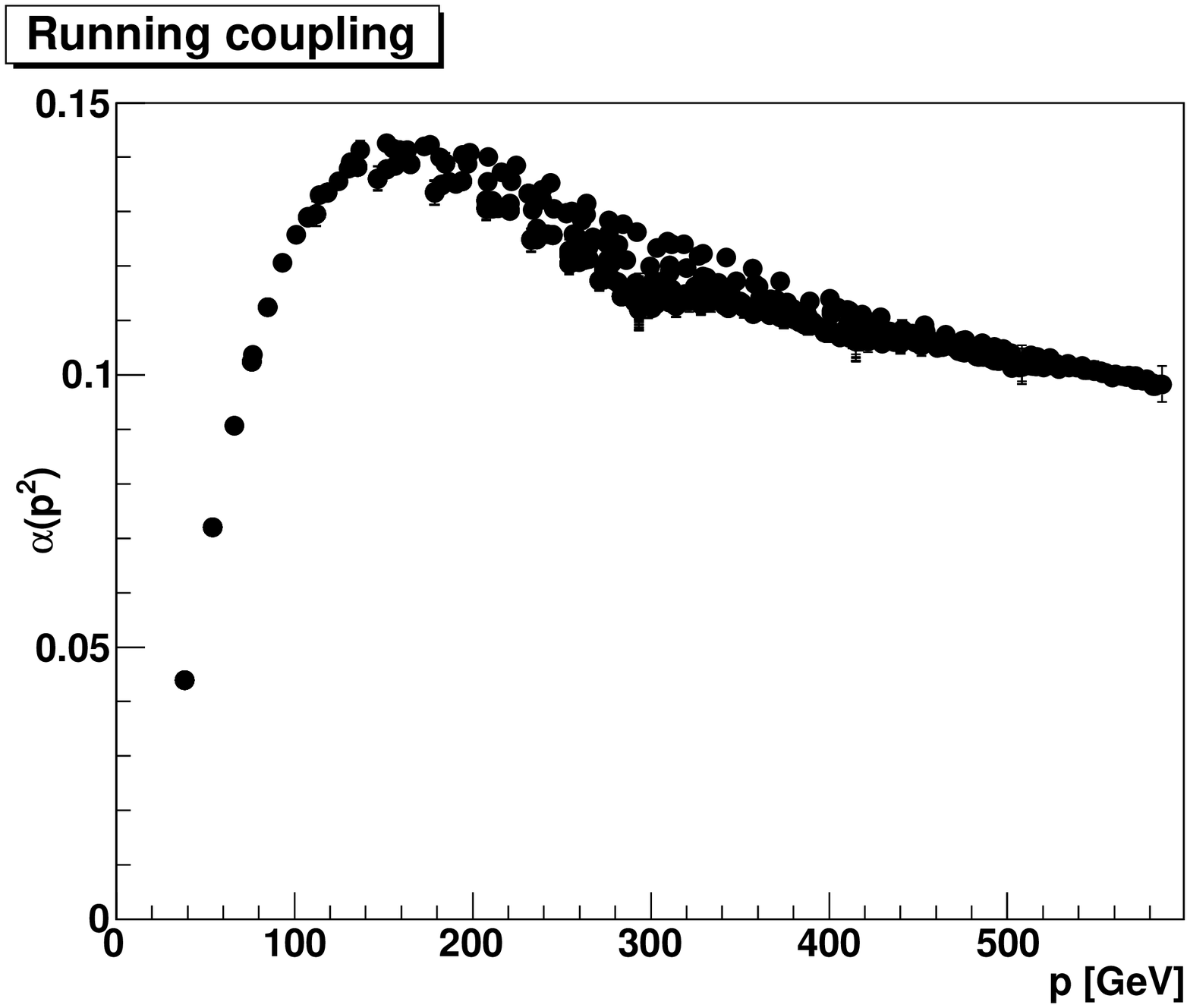}\includegraphics[width=0.5\linewidth]{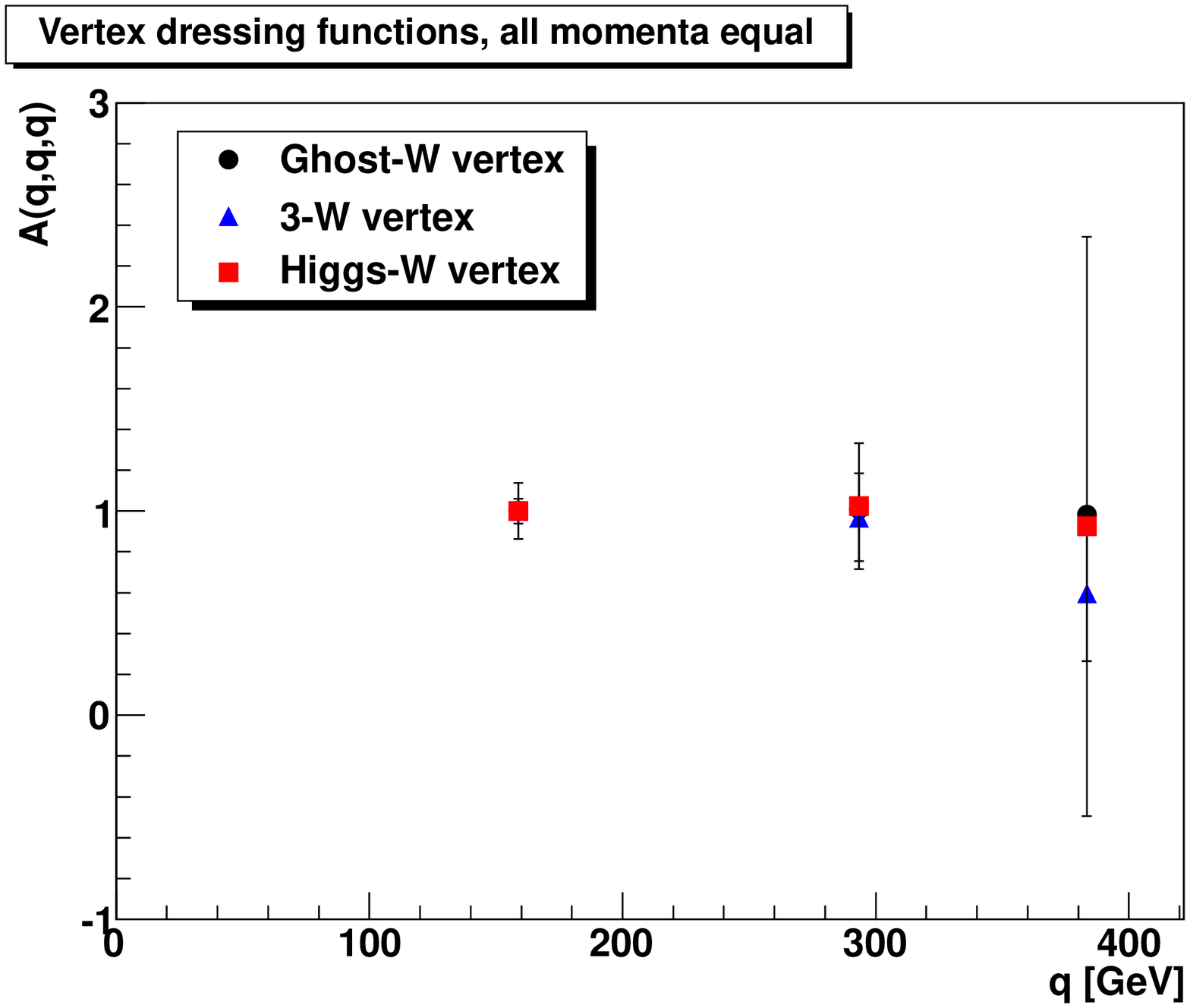}
\caption{\label{fig:3}The left panel shows the running weak fine structure constant, as determined from the two-point functions \cite{Maas:2010nc,vonSmekal:1997vx} on a 24$^4$ lattice with a lattice spacing of $(147$ GeV)$^{-1}$, see \cite{Maas:2010nc,Maas:2012tj} for details. The spreading of the points at mid-momentum is a lattice artifact \cite{Maas:2011se}. The right panel shows the dressing functions of the 3-point vertices with all momenta of equal size and incoming, from a $8^4$ lattice with the same lattice spacing, renormalized to one at 158 GeV. For the ghost-$W$ and Higgs-$W$ vertices, this is the only dressing function, while for the 3-$W$ vertex the projection on the tree-level vertex is shown, see \cite{Cucchieri:2006tf,Maas:unpublished3} for details.}
\end{figure}

The gauge interaction, however, is much more accessible, since it already appears at three-point level. Moreover, the running coupling can in Landau gauge be extracted already from the 2-point functions of the $W$ and the ghost alone \cite{vonSmekal:1997vx}. It is shown in the left panel of figure \ref{fig:3}. While it shows a slow decay towards higher momenta, it quickly vanishes in the infrared, just like in pure Yang-Mills theory \cite{Maas:2011se}.

Even more interesting are the full three-point vertices \cite{Maas:unpublished3}. In the current gauge, three of them exist. Two are entirely in the gauge sector, being the three-$W$ vertex and the $W$-ghost-vertex. In addition, there is the $W$-Higgs vertex. Concerning the latter, particular care is necessary. Since the Lagrangian \pref{action} exhibits, in addition to the local gauge symmetry, a unbroken global Higgs flavor symmetry \cite{Shifman:2012zz}, only such correlation functions are non-zero, which are invariant under both global color and flavor rotation. While this is the case for all propagators and the gauge vertices, the naive definition of the Higgs-$W$ vertex is not, and one has to relegate to a flavor-invariant one. Then the calculation of all vertices is straight-forward \cite{Maas:2011se,Maas:unpublished3}, since there are no disconnected contributions in the present gauge.

The results for the amputated and renormalized vertices at the symmetric point $|p|=|q|=|k|$ are shown in the right panel of figure \ref{fig:3}. All dressing functions do not deviate strongly from the tree-level result, at least within the relatively large errors. This is in agreement with the current experimental results \cite{pdg}. Much more precise results will be necessary to find if any relevant deviation from perturbation theory exists \cite{Maas:unpublished3}.

\section{Bound states: Towards Higgs sector spectroscopy}\label{sgi}

\begin{figure}
\includegraphics[width=\linewidth]{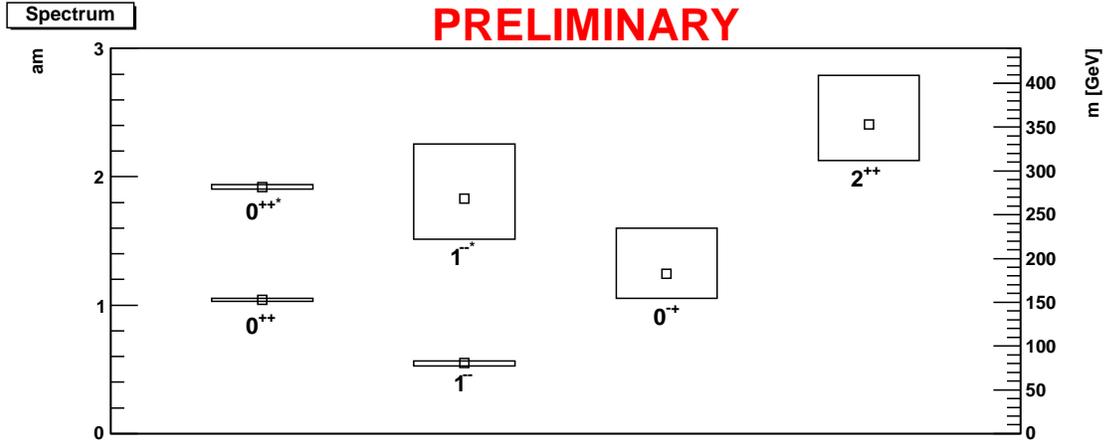}
\caption{\label{fig:4}The low-lying spectrum. See \cite{Maas:2012tj} for details on the calculation of the $0^{++}$ and $1^{--}$ states, which have been obtained on a 24$^4$ lattice. The data are not extrapolated to the continuum and infinite volume limit. The very preliminary results on the $0^{-+}$ and $2^{++}$ state are also not extrapolated and from a $8^4$ lattice, and contain only standard $W$-ball operators \cite{DeGrand:2006zz}. In both cases the lattice spacing is $a=(147$ GeV)$^{-1}$. Note that due to the flavor symmetry the $1^{--}$ is threefold degenerate. The square is the average value, while the box marks the statistical uncertainty.}
\end{figure}

In a non-Abelian gauge theory, only composite operators can be physical states \cite{Frohlich:1981yi}, and they can be interpreted as bound states, just as in QCD. This does not yet make a statement about the properties of these states, though in the current setup the lightest one will be necessarily stable, since without the rest of the standard model no decay channels are open. Such bound states have been calculated in the past for the theory described by \pref{action}, see e.\ g.\ \cite{Maas:2012tj,Langguth:1985dr,Evertz:1985fc,Philipsen:1996af}, and the lightest particle has been found to be massive. The low-lying spectrum of the theory is shown in figure \ref{fig:4}.

Already from these data an interesting observation can be made. The state with the same quantum numbers as the $W$, the $1^{--}$ state is threefold degenerate and has the same mass as the $W$, though the degeneracy is now a consequence of the flavor instead of the gauge symmetry. At the same time the state with the $0^{++}$ quantum numbers of the Higgs is found to be roughly at 150 GeV, and by a different choice of lattice parameters can likely be brought down to 125 GeV \cite{Maas:unpublished3}.

This similarity of the masses to the elementary particles is not coincidental. E.\ g.\ the $0^{++}$ state can be expanded in an appropriate gauge in the quantum fluctuations $\eta=\phi-\langle\phi\rangle$ of the Higgs field as \cite{Frohlich:1981yi}
\be
\langle\phi_i^+(x)\phi^i(x)\phi_j^+(y)\phi^j(y)\rangle\approx \langle\phi\rangle^4+\langle\phi\rangle^2(c'+\langle\phi^+(x)\phi(y)\rangle)+{\cal O}(\langle|\eta|\rangle)\label{expansion}.
\ee
\no which implies that to leading order the elementary Higgs and the $0^{++}$ bound state, made up out of 2 Higgs particles \cite{Maas:2012tj}, should have the same pole mass. A similar duality relation holds for the $W$ and the bound state in the $1^{--}$ channel. Thus the physical states are, to leading order in the quantum fluctuation of the Higgs field, indistinguishable from the bound states. Therefore, the physical observed resonances in experiments can, and should be \cite{Frohlich:1981yi,Maas:2012tj}, identified with the physical bound states, rather than with the gauge-dependent elementary degrees of freedom. Note that in a constitute picture the mass defect of these states is of the order of the constituent mass. These states are therefore deeply bound, relativistic states, and thus inaccessible to quantum-mechanical Schr\"odinger-type equations.

The remainder of the spectrum is then quite interesting. There are excited states, which may or may not be just scattering states, and states with more exotic quantum numbers. All of these states have no leading-order contribution in an expansion like \pref{expansion}. They are thus of higher order in the Higgs fluctuations. As a crude and very naive estimate, it can be expected that their production is thus suppressed at least by the ratio $\langle|\eta|\rangle/\langle\phi\rangle$, which for the present lattice setting is bound from above by 1\%. Thus, very crudely, at least 100 times the statistics will be necessary to identify these effects. Given that about 10 fb$^{-1}$ had been necessary to find the (possible) Higgs, this would imply about 1000 fb$^{-1}$, an amount of statistics the LHC may reach in the early 2020ies with a luminosity upgrade, or at the ILC. However, the 0$^{-+}$ state nearby to the $0^{++}$ state may already make itself notable earlier by parity-violating stray decays.

More importantly these states, if they are sufficiently stable to be detectable at all, can give rise to a standard model background to new physics searches. The 2$^{++}$ state, e.\ g., has the quantum numbers expected for a heavy Kaluza-Klein/Randall-Sundrum graviton. Understanding these states thoroughly is thus indispensable to make sure that they do not afflict these searches.

However, these investigations are yet at a very preliminary point. Though the existence of such states is a consequence of quantum field theory, their properties are not. These lattice simulations yet lack infinite-volume and continuum-extrapolation, a determination of decay properties, and, probably much more importantly, the effects of QED and fermions, which may influence such states quite significantly. Especially parity violations will be a serious challenge, but the combination of lattice and continuum methods, working so well for QCD \cite{Maas:2011se}, may provide an option. For this, the results from sections \ref{sgd2} and \ref{sgd3} will be very important. Finally, the determination of cross-sections non-perturbatively is highly complicated, especially for bound states. The use of effective low-energy theories may therefore be compulsory to finally arrive at quantitative predictions for experiments, beyond the mere prediction of the states. Nonetheless, if these states exist, they offer a quite intriguing manifestation of field theory, and a whole new arena for spectroscopy at experiments in the Higgs sector.

\bibliographystyle{bibstyle}
\bibliography{bib}

\end{document}